\documentclass[twocolumn,letter]{jpsj2}  
\title{Cooperative Phenomenon of Ferromagnetism and Unconventional Superconductivity in UGe$_2$: A $^{73}$Ge-NQR Study under Pressure}

\author{A. \textsc{Harada}$^{1}$, S. \textsc{Kawasaki}$^{1}$, H. \textsc{Kotegawa}$^{1}$\thanks{Present address: Department of Physics, Faculty of Science, Okayama University, Okayama 700-8530, Japan}, Y. \textsc{Kitaoka}$^{1}$, Y. \textsc{Haga}$^{2}$, E. \textsc{Yamamoto}$^{2}$, Y. \textsc{\=Onuki}$^{2,3}$, K. M. \textsc{Itoh}$^{4}$, E. E. \textsc{Haller}$^{5}$, and H. \textsc{Harima}$^{6}$}

\inst{$^{1}$Department of Materials Engineering Science, Graduate School of Engineering Science, Osaka University, Toyonaka, Osaka 560-8531, Japan \\
$^{2}$Advanced Science Research Center, Japan Atomic Energy Research Institute, Tokai, Ibaraki 319-1195, Japan\\
$^{3}$Department of Physics, Graduate School of Science, Osaka University, Toyonaka, Osaka 560-0043, Japan \\
$^{4}$Department of Applied Physics and Physico-Informatics,
Keio University, Yokohama 223-8522, Japan\\
$^{5}$Department of Materials Science and Engineering, University of California at Berkeley and Lawrence 
Berkeley National Laboratory, Berkeley, CA 94720, USA\\
$^{6}$Department of Physics, Faculty of Science, Kobe University, Nada, Kobe 657-8501, Japan}

\recdate{\today}

\abst{We report on a cooperative phenomenon of ferromagnetism and unconventional superconductivity (SC) in UGe$_2$ through the measurements of $^{73}$Ge nuclear-quadrupole-resonance (NQR) under pressure ($P$). 
The NQR spectra evidenced phase separation into ferromagnetic and paramagnetic phases in the vicinity of $P_c\sim 1.5$ GPa, pointing to a first-order transition. The measurements of nuclear-spin-lattice-relaxation-rate $1/T_1$ revealed that SC emerges under the background of ferromagnetism, but not of the paramagnetic phase.}

\kword{itinerant ferromagnetism, unconventional superconductivity, UGe$_2$, NQR under pressure, first-order phase transition}

\newpage
\begin{document}
\maketitle

Recently, superconductivity (SC) has been observed under a background of ferromagnetism (FM) in UGe$_2$ \cite{Saxena,Huxley} and URhGe \cite{Aoki}. Since FM and SC have been generally believed to be exclusive from each other, the coexsistence of FM and SC in these compounds is a great surprise. 
Figure 1(a) shows the pressure ($P$) versus temperature ($T$) phase diagram of UGe$_2$ established from various measurements \cite{Saxena,Huxley,Kobayashi,Haga,Tateiwa,Pfleiderer2}.
 The ferromagnetic transition temperature $T_{Curie}$ at ambient pressure ($P$ = 0) decreases monotonically from $T_{Curie}$= 52 K with increasing $P$ and seems to suddenly vanish around $P_c \sim 1.5$ GPa.
SC was discovered in the range of $P$ = 1 to 1.5 GPa, exhibiting the highest transition temperature $T_{sc}\sim 0.7$ K at $P_x\sim 1.2$ GPa.
Upon cooling below $T_x$, ferromagnetic moments increase markedly \cite{Saxena,Huxley,Tateiwa}. The high-$P$-$T$ and the low-$P$-$T$ phases are denoted as FM1 and FM2 respectively, as indicated in the phase diagram.
At $P_x$, the first-order transition from FM2 to FM1 emerges as a function of $P$ and at $T_x$ as a function of $T$ in the case of $P < P_x$. Here, $P_x$ is a terminal point of the first-order transition \cite{Pfleiderer2,Kotegawa3}. 

By contrast, it should be noted that both SC and FM1 are simultaneously suppressed at a ferromagnetic critical pressure $P_c$. This suggests that SC and FM in UGe$_2$ are cooperative phenomena. Indeed, the $^{73}$Ge-NQR measurements 
have revealed that both FM1 and FM2 coexist uniformly with the unconventional SC with a line node-gap in the vicinity of $P_x$ \cite{Kotegawa1,Kotegawa2,Kotegawa3}. 
It is surprising that the 5$f$ electrons of uranium contribute to the onset of the uniform coexistent phase of FM and SC. This phenomenon is difficult to understand in terms of the spin-singlet pairing framework for the Cooper pairs. Therefore, this fact suggests that SC in UGe$_2$ may be in a spin-triplet pairing state under the background of FM.
In order to address the intimate interplay between FM and SC in UGe$_2$ near $P_c\sim 1.5$ GPa where FM seems to collapse, we here report on the microscopic characteristics of FM and SC near and over $P_c$ revealed by the measurements of nuclear-quadrupole-resonance (NQR) of enriched $^{73}$Ge under $P$ at zero  magnetic field ($H=0$) . 
\begin{figure}[h]
\centering
\includegraphics[width=7.5cm]{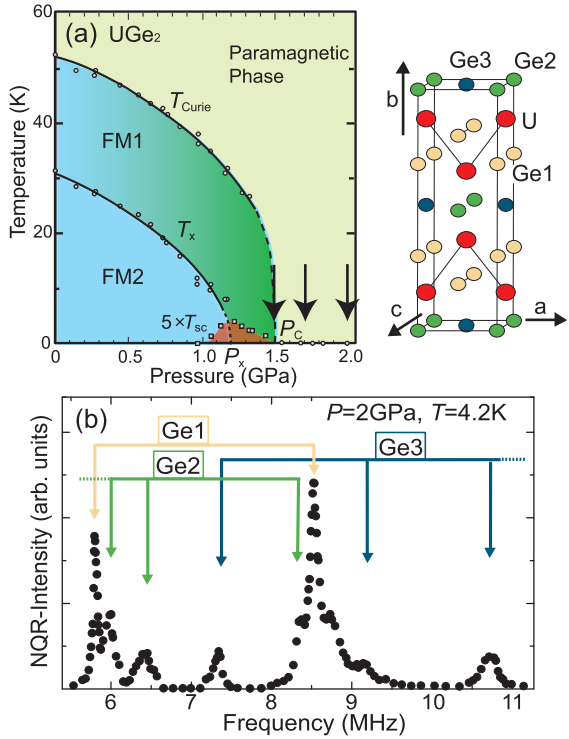}
\caption[]{(Color online) (a) The pressure ($P$) versus temperature ($T$) phase diagram of UGe$_2$ \cite{Haga,Tateiwa,Kobayashi}. The arrows denote the values of 
$P$ where the present NQR measurements have been performed. (b) The NQR spectra for the paramagnetic state at $T= 4.2$ K and $P=$ 2 GPa. 
It reveals a structure consisting of significantly separated peaks associated with three inequivalent Ge sites in one unit cell.}
\end{figure}
A polycrystalline sample enriched with $^{73}$Ge was prepared and crushed into powder to allow a maximal penetration of oscillating
 magnetic field into the sample. The NQR experiment was performed by the conventional spin-echo method at $H=0$ in the frequency ($f$) range of 5 to 12 MHz at $P$ = 1.5, 1.7 and 2.0 GPa.
Hydrostatic pressure was applied by utilizing a NiCrAl-BeCu piston-cylinder cell filled with Daphne oil (7373) as a pressure-transmitting medium. In order to calibrate each value of $P$ at low temperatures, the $P$ dependence of $T_c$ of Sn metal was measured by resistivity measurement. 
A $^{3}$He-$^{4}$He dilution refrigerator was used to reach the lowest temperature, 40 mK. 
Figure 1(b) shows the NQR spectra for the paramagnetic phase (PM)  at $T = 4.2$ K and $P = 2.0$ GPa. They reveal a 
structure consisting of significantly separated peaks associated with three inequivalent Ge sites in one unit cell (see Figure 1).
The number of Ge1 in one unit cell is two times as large as the number of either Ge2 or Ge3 in one unit cell. The Ge1 site is closely located along the uranium (U)-zigzag chain, while the other two sites Ge2 and Ge3 are located out of 
this zigzag chain. The values of NQR parameters at the Ge1, Ge2 and Ge3 sites were estimated on the basis of the band calculation by one of authors (H. Harima) \cite{Kotegawa3}.

\begin{figure}[h]
\centering
\includegraphics[width=7.6cm]{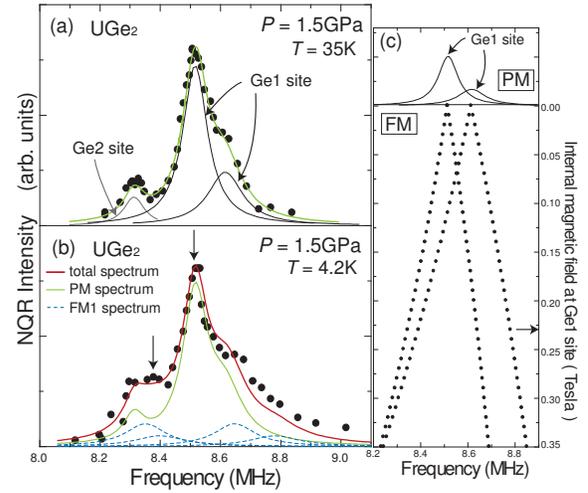}
\caption[]{(Color online) (a) The $^{73}$Ge spectra at $T=35$ K and $P_c\sim$ 1.5 GPa in the paramagnetic phase (PM). Over the range of $f$
 = 8.2 to 8.8 MHz there are two NQR spectra for the Ge1 site (middle and right) and the one NQR spectrum for the Ge2 site (left). 
 (b) The $^{73}$Ge spectra at $T=4.2$ K where the phase separation takes place. It is well simulated by the respective spectra arising from PM and FM1 indicated by the lower solid and dotted lines, allowing us to estimate the ratio of PM to FM1 to be 7:3 and $H_{int}\sim$ 0.2 T at the Ge1 site due to the onset of FM1.
(c) The variation (dotted lines) in frequencies for the two peaks in the spectra for the Ge1 site in PM, which are separated as $H_{int}$ increases in association with the onset of  FM1. The spectra (b) that reveal the phase separation into PM and FM1, are reproduced by assuming $H_{int}$= 0.23 T at 4.2 K for FM1, as marked by the arrow in (c).
}
\end{figure} 

Figure 2(a) indicates the spectra for the PM at $T=35$ K and $P_c\sim 1.5$ GPa where FM1 collapses. Over the range of $f$= 8.2 to 8.8 MHz there are two NQR spectra for the Ge1 site (middle and right) and one NQR spectrum for the Ge2 site (left). 
Note that the spectra at the Ge1 site almost overlap due to the asymmetry parameter $\eta\sim 1$.  As temperature decreases below 10 K, the spectral shape becomes slightly broadened as seen for the spectra at $T=4.2$ K in Figure 2(b), although the overall shape remains unchanged, suggesting that the system remains mostly in PM. This result shows that PM and FM1 are separated at $P_c\sim$ 1.5 GPa.  By contrast, the NQR spectra at $P=1.7$ and 2.0 GPa beyond $P_c$ reveal that the system remains in PM down to 0.5 K.
Here we focus on the change in NQR spectra caused by the small size of $H_{int}$ that is induced at the Ge1 site due to the onset of FM1. This is because the Ge1 site is located near the uranium zigzag chain responsible for FM1 and that the number of the Ge1 site is twice larger than that of the Ge2 and Ge3 sites in one unit cell. 
The dotted lines in Figure 2(c) indicate the variation in frequencies for the two peaks in the spectra at the Ge1 site in PM, which are separated as $H_{int}$ increases for FM1.
The spectra in Figure 2(b) are reproduced by assuming $H_{int}$= 0.23 T at $T=4.2$ K for FM1, as marked by the arrow in Figure 2(c) and also assuming the ratio of PM to FM1 of 7:3. This means that the phase separation into PM and FM1 takes place.
\begin{figure}[htbp]
\centering
\includegraphics[width=6.2cm]{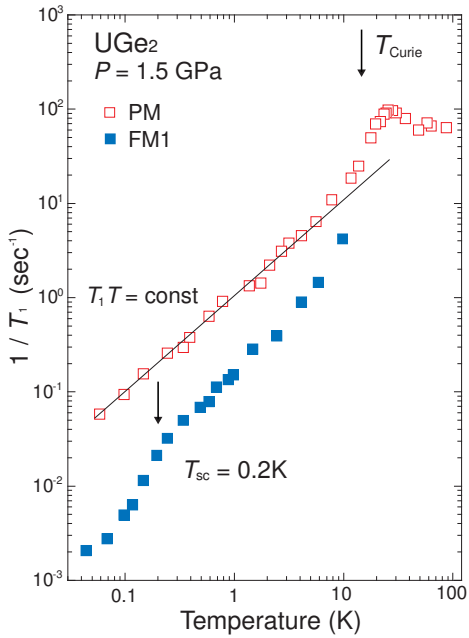}
\caption[]{(Color online) The $T$ dependence of $1/T_1$ for PM (open squares) and FM1 (solid squares). The long components in $1/T_1$ for FM1 indicate that SC sets in at $T_{sc}\sim 0.2$ K, but the short components for PM do not.}
\end{figure}
Thus, the first-order transition from FM1 to PM is evidenced around $P_c\sim$ 1.5 GPa from a microscopic point of view. It is noteworthy that in the previous work, both FM1 and FM2 coexist in a certain $T$ range lower than $T_x$ just below $P_x\sim 1.2$ GPa as well \cite{Kotegawa3}. 

Figure 3 shows the $T$ dependence of $1/T_1$ at $P_c\sim$ 1.5 GPa and $H$= 0. As temperature decreases below 10 K, the two components in $T_1$ appear in association with the onset of FM1 because the phase separation into PM and FM1 is evidenced from the NQR spectra below $T\sim 15$ K. The $T_1$'s for FM1 and PM are measured at the respective frequencies 8.37 MHz and 8.51 MHz, which are marked by the arrows in Figure 2(b). Note that the $T_1$ at 8.51 MHz is uniquely determined by the theoretical recovery curve of NQR relaxation at the Ge1 site in the case of $\eta=1$\cite{Kotegawa3}. On the other hand, the careful analysis of $T_1$ at 8.37 MHz is required because the spectra of FM1 overlap with those arising from  the Ge1 and Ge2 sites remaining in the PM. Noting that the NQR intensity  in the spectrum simulated as arising from the Ge2 in Figure 2(a) is very small at 8.37 MHz and that the intensity ratio of FM1 to PM for the Ge1 site is estimated to be 6:4 at 8.37 MHz, the respective values of $T_1(FM1)$ and $T_1(PM)$ for FM1 and PM are reasonably deduced from fitting a measured recovery curve of nuclear magnetization, $m_{obs}$, to a theoretical one, $A\times m_{Ge1,4\nu_Q}[t/T_1(FM1)]+B\times m_{Ge1,4\nu_Q}[t/T_1(PM)]$ (with $A=0.6$ and $B=0.4$)
. Here, $m_{obs}=\frac{M(\infty)-M(t)}{M(\infty)}$, where ${M(\infty)}$ and ${M(t)}$  are the respective values at the thermal equilibrium state and at a time $t$ after saturation pulses. The theoretical relaxation curve for $4\nu_Q$ transition is given by $m_{Ge1,4\nu_Q}[t/T_1]=0.075{\rm exp}(-\frac{3t}{T_1})+0.347{\rm exp}(-\frac{7.5t}{T_1})+0.425{\rm exp}(-\frac{16.5t}{T_1})+0.153{\rm exp}(-\frac{27.6t}{T_1})$. 
A notable result is that the $T_1(FM1)$ (solid squares) associated with FM1 decreases below $T\sim$ 0.2 K. This decrease in  $1/T_1(FM1)$ may be ascribed to a superconducting transition emerging for FM1. This is because the anomalies related to the onset of SC at $T_{sc}\sim 0.2$ K were actually observed  near $P_c$ in the other measurements \cite{Saxena,Huxley,Kobayashi,Tateiwa2}. Remarkably, no trace for SC is seen in the $1/T_1(PM)$ (open squares) for PM, indicating that the SC in UGe$_2$ sets in under the background of FM. 
 
Figure 4(a) and 4(b) indicate the $T$ dependence of $1/T_1T$ for the FM1 at $P_x\sim 1.2$ GPa and $P_c\sim 1.5$ GPa and for the PM at 1.5, 1.7 and 2.0 GPa, respectively. At low temperatures, the value of $T_1T$ remains constant. It is remarkable that the $P$ dependence of $(1/T_1T)^{1/2}$ is scaled to that of the $T$-linear coefficient in heat capacity $\gamma$ (open squares), as presented in Figure 4(c). 
\begin{figure}[htbp]
\centering
\includegraphics[width=7.2cm]{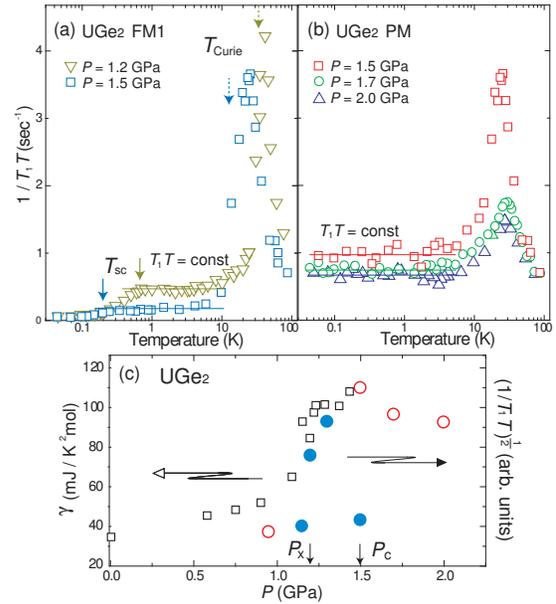}
\caption[]{(Color online) (a) The $T$ dependencies of $1/T_1T$ for FM1 at $P_x\sim 1.2$ GPa and $P_c\sim 1.5$ GPa, and (b) for PM at $P=1.5$, 1.7 and 2.0 GPa. (c) The $P$ dependence of ($1/T_1T)^{1/2}$, which is scaled to that of the $T$-linear coefficient at heat capacity $\gamma$ (open squares). The data of $\gamma$ are taken from ref. 6.  Note that $(1/T_1T)^{1/2}$ is indicated by solid circles for FM1 and FM2 where SC sets in and by open circles for FM2 and PM where SC collapses.}
\end{figure}
Here the data of $\gamma$ are taken from ref. 6. This coincidence between the $P$ dependencies of $(1/T_1T)^{1/2}$ and $\gamma$ points to the intimate $P$-derived variation of the effective density of states (DOS) at the Fermi level.  Note that the DOS for FM1 and FM2 increases suddenly around $P_x\sim$ 1.2 GPa where the first-order transition from FM2 to FM1 occurs, whereas it decreases at $P_c\sim 1.5$ GPa, exhibiting a sharp peak around $P_x\sim 1.2$ GPa. Note that the peak in $\gamma$ at $P_c\sim 1.5$ GPa arises from PM separated from FM1.  Interestingly, the DOS at $P_c\sim 1.5$ GPa for FM1 is comparable to that for FM2 at 
$P=1.15$ GPa, and then the respective SC sets in at $T_{sc}\sim 0.2$ and 0.35 K for FM1 and FM2. In the vicinity of $P_x\sim 1.2$ GPa, where the DOS for FM1 has its peak, the SC undergoes the highest $T_{sc},\sim$ 0.7 K \cite{Pfleiderer2,Kotegawa3}. It is the peak in DOS around $P_x\sim$ 1.2 GPa that leads to be highest $T_{sc}$.
We remark, however, that a larger DOS for PM than for FM1 at $P_c\sim 1.5$ GPa does not always lead to the onset of SC. It is evident that the onset of SC takes place under the background of FM in the range of $P$= 1.15 to 1.5 GPa. 
As can be seen in Figure 4(a), $1/T_1T$ at $P_x\sim 1.2$ GPa has a peak at $T_{Curie}=31$ K due to the critical slowing down of spin fluctuations associated with the FM1 ordering.
However, as seen in Figure 4(b), $1/T_1T$ at $P_c\sim 1.5$ GPa also exhibits a peak at around $T_p\sim 25$ K. This temperature is higher than the phase-separation temperature $T\sim$ 15 K. Furthermore, 
even at $P$= 1.7 and 2.0 GPa, where FM1 completely collapses, it is unusual that a broad peak in $1/T_1T$ is observed  at around $T_p\sim 30$ K for PM.
In the paramagnetic state well above $T_{Curie}$ and $T_p$, $1/T_1T$ is expected to be expressed by $(T_1T)\propto 1/\chi(T)\propto (T-T_{Curie})$, which is predicted by the self-consistent renormalization (SCR) theory for weakly itinerant ferromagnets \cite{Moriya}. Here, $\chi(T)$ is the uniform susceptibility. Actually, the $1/T_1T$ data at $P_x\sim 1.2$ GPa and $P_c\sim 1.5$ GPa follow the relation $(T_1T)\propto (T-T_{Curie})$, with $T_{Curie}=31$ K and 11 K, respectively, as shown in Figure 5. Note here that $T_{Curie}=11$ K at $P_c\sim 1.5$ GPa is close to $T\sim$ 15 K, but it is significantly lower than $T_p\sim 25$ K, where the unexpected peak in $1/T_1T$ is observed, as seen in Fig.4(a).
At $P$= 1.7 and 2.0 GPa, where FM1 completely collapses, $(T_1T)\propto 1/\chi(T)\propto (T-\theta)$ seems to be valid as predicted for nearly ferromagnetic metals, with $\theta\sim -3$ and $-11$ K, respectively.
If an isotropic spin-fluctuation regime was dominant in these nearly ferromagnetic systems, then this relation should be valid down to low $T$. For UGe$_2$ at $P$= 1.7 and 2.0 GPa, however, the broad unexpected peak emerges around $T_p\sim 30$ K, as seen in Figure 4(b).
It is suggested that this suppression of spin fluctuations below $T_p$ may be caused by anisotropic spin fluctuations of $5f$ electrons because of the Ising-like uniaxial magnetic properties in UGe$_2$, which are reported in the literature \cite{Huxley2}. If such anisotropic character in the dynamical susceptibility of spin fluctuations reduces the transversal components of hyperfine fields at the Ge sites, the nuclear relaxation rate $1/T_1$ might be suppressed.    
This anisotropic character of spin fluctuations in PM may be relevant with the metamagnetic increase in magnetization upon increasing $H$ \cite{Haga,Pfleiderer2}.
This is also consistent with the fact that the magnetization in either FM1 or FM2 exhibits the Ising uniaxial anisotropy along the a-axis. Interestingly, in FM1 the application of a magnetic field makes the upper critical field jump rapidly and $T_{sc}$ in zero field seems to be enhanced, whereas applying pressure decreases $T_{sc}$. 
On the other hand, even though PM undergoes the metamagnetic increase in magnetization and enters the field-induced FM1, but supeconductivity does not take place near $P_c\sim 1.5$ GPa. Noting that the SC in UGe$_2$ exhibits the highest $T_{sc}$ at the $P_x$ that is the termination point of first-order transition from FM2 to FM1, the longitudinal fluctuations of ferromagnetic polarization emerging in either FM1 or FM2 are thought to play a role in mediating the Cooper pairs in UGe$_2$ because the stiffness of Ising-like ferromagnetic moments becomes soft at $P_x$. 

In conclusion, the $^{73}$Ge-NQR measurements on UGe$_2$  
 have revealed that FM1 and PM are separated at $P_c\sim 1.5$ GPa.
\begin{figure}[h]
\centering
\includegraphics[width=7cm]{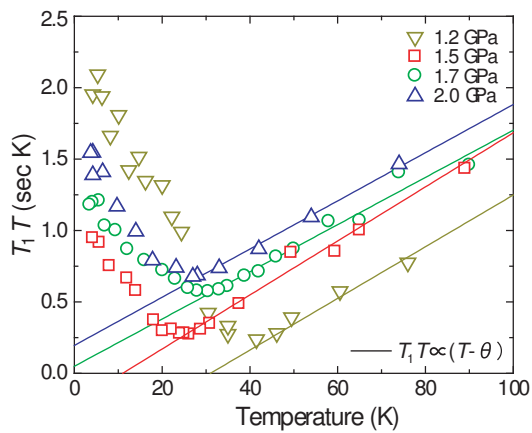}
\caption[]{(Color online) The $T$ dependencies of $T_1T$ at $P$ = 1.2 GPa (open inverted triangles), 1.5 GPa (open squares), 1.7 GPa (open circles) and 2.0 GPa (open triangles). As predicted from the self-consistent renormalization (SCR) theory for weakly and nearly itinerant ferromagnetic metals, the relation of $(T_1T)\propto 1/\chi(T)\propto (T-T_{Curie})$ is valid for the data at $P_x\sim 1.2$ GPa with $T_{Curie}=31$ K. This figure also shows the FM1 at $P_c\sim 1.5$ GPa with $T_{Curie}=11$ K.  At respective $P=1.7$ and 2.0 GPa where FM1 collapses, $(T_1T)\propto 1/\chi(T)\propto (T-\theta)$ is valid, with $\theta\sim -3$ and $-11$, respectively, K as expected for nearly ferromagnetic metals.}
\end{figure}
A notable result is that the SC at $T_{sc}\sim$ 0.2 K emerges only under the background of FM1, but not of PM near $P_c\sim 1.5$ GPa. Furthermore, the results at pressures exceeding $P_c$ have suggested that the spin fluctuations at low $T$ possess the Ising-like character even in the $P$-induced PM. This anisotropic character of spin fluctuations in PM may be relevant to the metamagnetic increase in magnetization upon increasing $H$.
Noting that the field-induced FM1 that is caused by the metamagnetic increase in magnetization does not reveal supeconductivity, it is expected that the onset of SC in UGe$_2$ is mediated by the longitudinal fluctuations of ferromagnetic polarization in FM, which become significantly soft near $P_x$ where the first-order transition from FM2 to FM1 occurs.


This work was supported by a Grant-in-Aid for Creative Scientific Researchi15GS0213), MEXT, and the 21st Century COE Program supported by Japan Society for the Promotion of Science.
One of the authors (S. K.) has been supported by a Research Fellowship of the Japan Society for the 
Promotion of Science for Young Scientists.


\end{document}